\documentclass[twocolumn, pra, showpacs,superscriptaddress]{revtex4}
\usepackage{graphicx}
\usepackage{dcolumn}
\usepackage{bm}
\usepackage{amsmath}

\begin{document}

\title{One-dimensional fermionic gases with attractive $p$-wave
interaction in a hard-wall trap}
\author{Yajiang Hao}
\affiliation {Beijing National Laboratory for Condensed Matter
Physics, Institute of Physics, Chinese Academy of Sciences, Beijing
100080, P. R. China}
\author{Yunbo Zhang}
\affiliation{Department of Physics and Institute of Theoretical
Physics, Shanxi University, Taiyuan 030006, P. R. China}
\author{Shu Chen}
\email{schen@aphy.iphy.ac.cn} \affiliation{Beijing National
Laboratory for Condensed Matter Physics, Institute of Physics,
Chinese Academy of Sciences, Beijing 100080, P. R. China}
\date{Original August 27, 2007; Revised \today}

\begin{abstract}
We investigate the ground state of the one-dimensional fermionic
system enclosed in a hard-wall trap with attractive contact
$p$-wave interactions. Based on the Bethe ansatz method, the
explicit wave function is derived by numerically solving the Bethe
ansatz equations for the full physical regimes ($-\infty \leq
c_F\leq 0$). With the exact wave function some quantities which
are important in many-body physics are obtained, including the
one-body density matrix and the momentum distribution of the
ground state for finite system. It is shown that the shell
structure of the density profiles disappears with the increase of
the interaction and in the fermionic Tonks-Girardeau (FTG) limit
the density distribution shows the same behavior as that of an
ideal Bose gas. However the one-body density matrix and the
momentum distribution exhibit completely different structures
compared with their bosonic counterparts.
\end{abstract}
\maketitle



\narrowtext
\section{Introduction}
Experimental realization of trapped one-dimensional (1D) cold atom
systems \cite{gorlitz,Paredes,Toshiya,esslinger} are triggering more
and more theoretical efforts to study the 1D many-body physics
beyond the mean-field theory. For the ultracold quantum gases
tightly confined in waveguides, the dynamics are effectively
described by a 1D model due to the radial degrees of freedom are
frozen \cite{Olshanii,Blume}. Further, the ability of tuning the
effective 1D interactions by Feshbach resonance allows experimental
access to the very strongly interacting regime where correlation
effects are greatly enhanced \cite
{Petrov,Dunjko,Chen,Olshanii2,ohberg}. In the limit of the
Tonks-Girardeau (TG) \cite{Girardeau2} gas with effective coupling
constant $g_{1D}\rightarrow\infty$, the many-body problem of a TG
gas can be mapped to that of a free Fermi gas by the Bose-Fermi
mapping, which has been verified by two experimental groups
\cite{Paredes,Toshiya}. This Bose-Fermi duality was generalized to
show the equivalence between a 1D fermionic system and a bosonic one
with the reversed role of strong and weak couplings \cite{Cheon}.
Recently, the exact ground state of the fermionic TG (FTG) gas,
defined as a 1D spin-polarized fermionic gas with infinitely strong
attractive $p$-wave interactions, has been determined by inversely
Fermi-Bose mapping to the ideal Bose gas
\cite{Girardeau05,Girardeau06,Girardeau04,Granger}.

A key experimental challenge is to obtain superfluidity with pairs
in nonzero orbital angular momentum states by using $p$-wave, or
maybe even $d$-wave Feshbach resonances. In general, the $p$-wave
interaction is very weak comparing with the $s$ -wave interaction.
However, for a spin-polarized fermionic gas, the $s$-wave scattering
is forbidden due to the Pauli exclusion principle and thus the $p$
-wave interaction is dominant. Furthermore, the $p$-wave interaction
can be greatly enhanced by the Feshbach resonances \cite
{Regal,Schunck,JingZhang,Gunter} and using a $p$-wave Feshbach
resonance between $^{40}K$ atoms Jin's group at JILA have
successfully produced and detected molecules with lifetimes on the
order of milliseconds on both the BEC and the BCS side of the
resonance \cite{Gaebler}. For a 1D gas, the additional confinement
induced resonance permits one to tune the 1D effective interaction
via a 3D Feshbach resonance \cite {Olshanii,Blume,Gunter}.

In this paper, we report on a detailed study of the 1D Fermi gases
in the infinitely deep square potential well. We will show that the
model of fermionic gases with attractive $p$-wave interactions in
such a one-dimensional hard-wall trap is exactly solvable by the
Bethe-ansatz method. The experimental efforts in trapping ultracold
gases near micro-fabricated surfaces, the so-called "atom chips"
\cite{Hansel,ZimmermannRMP}, and various innovative features in
designing the optical box trap \cite{Raizen}, are specifically aimed
at studying the surprisingly rich variety of physical regimes
predicted for the 1D Bose gas and have stimulated many theoretical
studies on the physics in a box trap \cite{Hao,Niu,Muga}. Different
from the harmonic trap, the interacting model in a hard-wall trap is
integrable and thus could provide us some exact pictures for
understanding the trapped many-body systems. So far, there has been
a growing interest in the exactly solved models in the hard-wall
trap \cite{Gaudin,Guan,Hao}, but most of them focus on the Bose gas
and the Fermi gas with odd-wave interactions is not addressed. While
the theoretical understanding of the correlation effect of bosonic
system has been investigated extensively \cite
{Olshanii,Olshanii2,Petrov,Dunjko,Chen,ohberg}, the fermionic system
is not well understood except in the so-called FTG limit
\cite{Girardeau05,Girardeau06,del Campo,Girardeau04,Granger}.

\section{formulation of the model and its exact solution}
We consider an $N$-particle system with finite, attractive $p$-wave
interaction in a one-dimensional box of length $L$, which obviously
fills the gap between the FTG limit and free Fermions. The
Schr\"{o}dinger equation can be formulated as
\begin{equation}
\left[ -\sum_{i=1}^N\frac{\hbar ^2}{2m}\frac{\partial ^2}{\partial x_i^2}%
+\sum_{1\leq i<j\leq N}V\left( x_i-x_j\right) \right] \Psi =E\Psi ,
\end{equation}
where $V\left( x_i-x_j\right) $ is the pseudo-potential describing
the $p$ -wave scattering. It has been shown that the $p$-wave
scattering of two spin-polarized fermions in a tightly confined
waveguide can be well described by the contact condition
\cite{Blume,Girardeau04}
\begin{eqnarray}
\Psi _F\left( x_i-x_j=0^{+}\right)  &=&-\Psi _F\left( x_i-x_j=0^{-}\right)
\nonumber \\
&=&-a_{1D}^F\frac \partial {\partial x}\Psi _F\left( x_i=x_j\pm 0\right) ,
\end{eqnarray}
where
\begin{equation}
a_{1D}^F=\frac{3a_p^3}{l_{\perp }^2}\left[ 1+\frac{3\zeta (3/2)}{2\sqrt{2}%
\pi }\left( \frac{a_p}{l_{\perp }}\right) ^3\right] ^{-1}
\end{equation}
is the effective 1D scattering length with $a_p$ the $p$-wave scattering
length and $l_{\perp }=\sqrt{\hbar /m\omega _{\perp }}$ the transverse
oscillator length \cite{Girardeau04}. The contact condition can be
reproduced by using the following pseudo-potential \cite{Grosse,Girardeau04}
\begin{equation}
V\left( x\right) =-\frac{2\hbar ^2a_{1D}^F}m\frac \partial {\partial
x}\delta \left( x\right) \frac \partial {\partial x}
\end{equation}
where $x=x_i-x_j$ and $\partial _x=(\partial _{x_i}-\partial
_{x_j})/2.$  The scattering length can be tuned readily from $0$ to
$-\infty $ by sweeping an external magnetic field - the Feshbach
resonance, or by changing the geometry of the trapping potential -
the confinement induced resonance and in this paper the full
physical regimes $-\infty <a_{1D}^F<0$ will be studied. Similar to
the case of Bose gas, the important parameter characterizing the
different physical regimes of the 1D Fermi gas is $\gamma
=mg_{1D}^F\rho /\hbar ^2$, where $g_{1D}^F=-2\hbar ^2a_{1D}^F/m$ and
$\rho =N/L$.

A standard rescaling procedure brings the Schr\"{o}dinger equation
into a dimensionless one (for simplicity we keep the original
notations)
\[
H\Psi \left( x_1,\cdots ,x_N\right) =E\Psi \left( x_1,\cdots ,x_N\right)
\]
with
\[
H=-\sum_{i=1}^N\frac{\partial ^2}{\partial x_i^2}-2c_F\sum_{i<j}(\frac
\partial {\partial x_i}-\frac \partial {\partial x_j})\delta (x_i-x_j)(\frac
\partial {\partial x_i}-\frac \partial {\partial x_j}),
\]
where in the dimensionless interaction constant
$2c_F=\sqrt{2m/\hbar^2} a_{1D}^F$ we intentionally keep the factor 2
in accordance with the bosonic case \cite{Hao}. The wave function
takes the general form
\begin{eqnarray}
\Psi \left( x_1,\cdots ,x_N\right)  &=&\sum_Q \theta \left(
x_{q_N}-x_{q_{N-1}}\right) \cdots \theta
\left( x_{q_2}-x_{q_1}\right)  \nonumber \\
&&\times  \varphi_Q\left( x_{q_1},x_{q_2},\cdots ,x_{q_N}\right) ,
\label{WF}
\end{eqnarray}
where we have used $Q$ to label the region $0\leq x_{q_1}\leq
x_{q_2}\leq \cdots \leq x_{q_N}\leq L$. The wave function of
Fermions should follow the antisymmetry of exchange, so our model
is simplified into the solution of
\begin{equation}
H\varphi_Q\left( x_{q_1},\cdots ,x_{q_N}\right) =E\varphi_Q\left(
x_{q_1},\cdots ,x_{q_N}\right)   \label{sch}
\end{equation}
in the region $Q$ with the open boundary condition
\[
\varphi_Q\left( 0,x_{q_2},\cdots ,x_{q_N}\right) =\varphi_Q\left(
x_{q_1},x_{q_2},\cdots ,L\right) =0.
\]

Using the Bethe ansatz method we obtain the wavefunction
parameterized by the set of quantum number ${k_1,k_2,\cdots ,k_N}$
known as quasi-momenta or rapidities \cite{Hao}
\begin{eqnarray*}
&&\varphi_Q\left( x_{q_1},x_{q_2},\cdots ,x_{q_N}\right) \\
&=&(-1)^Q\sum_P (-1)^PA_p\exp \left[ i\left( \sum_{l<j}^{N-1}\omega
_{p_jp_l}\right) \right] \sin \left( k_{p_1}x_{q_1}\right) \\
&&\times \prod_{1<j<N}\sin \left( k_{p_j}x_{q_j}-\sum_{l<j}\omega
_{p_lp_j}\right) \\
&&\times \exp \left( ik_{p_N}L\right) \sin \left( k_{p_N}\left(
L-x_{q_N}\right) \right)
\end{eqnarray*}
with $\omega _{ab}=\arctan \frac{1/c_F}{k_a-k_b}-\arctan
\frac{1/c_F}{k_b+k_a }$ and $A_{p_1p_2...p_N}=\prod_{j<l}^N\left(
ik_{p_l}-ik_{p_j}-1/c_F\right) \left(
ik_{p_l}+ik_{p_j}-1/c_F\right)$. Here  $(-1)^Q=\pm 1$ and
$(-1)^P=\pm 1$ denote sign factors associated with even/odd
permutations of $Q=(q_1,q_2,\cdots,q_N)$ and
$P=(p_1,p_2,\cdots,p_N)$, respectively. The quasi-momenta
$k_1,k_2,\cdots ,k_N$ can be easily determined from the Bethe
ansatz equations
\[
\exp \left( i2k_jL\right) =\prod_{l=1\left( \neq j\right) }^N\frac{%
k_j-k_l-i/c_F}{k_j-k_l+i/c_F}\frac{k_j+k_l-i/c_F}{k_j+k_l+i/c_F}
\label{BA}
\]
with $j=1,2,\cdots ,N$. These quasi-momenta lead us immediately to
important physical quantities for our system. For example, the
energy eigenvalue is given by
\begin{equation}
E=\sum_{j=1}^Nk_j^2 \label{Energy}
\end{equation}
and the total momentum by $K=\sum _{j=1}^Nk_j$. It is clear that
these Bethe ansatz equations are the same as those in the case of
Bose gas if we simply make a substitution $c=-1/c_F$ \cite{Hao}.
In the regime $Q$, there is one-to-one correspondence between the
quasi-momentum solution of attractive $p$-wave Fermi gas and that
of repulsive Bose gas, but the total wave functions $\Psi
(x_1,x_2,\cdots ,x_N)$ take different forms because of their
distinct exchange symmetries. This difference can be easily seen
from the one-body density matrix and the momentum distribution.
Taking the logarithm of Bethe ansatz equations, we have
\begin{equation}
k_jL=n_j\pi +\sum_{l=1\left( \neq j\right) }^N\left( \arctan \frac{1/c_F}{%
k_l-k_j}-\arctan \frac{1/c_F}{k_j+k_l}\right)  \label{BAE}
\end{equation}
For the ground state the set of integer $n_j=1$ ($1\leq j\leq N$).
For simplicity, in the following evaluation the length $L$ will be
taken as one. It is obvious that the solutions of $k_j$ is only
relevant to $c_F/L$ for different $L$. The subsequent procedure is
standard. By numerically solving the transcendental equations
eqs.(\ref{BAE}), we obtain the quasi-momenta ${k_j}$ and thus the
ground state wave function. In principle, all necessary
information about the system can be inferred, including the
one-body density matrix, the momentum distribution, and the
excitation spectrum. Furthermore, one can apply the thermodynamic
formalism for dealing with the one dimensional interacting systems
developed by Yang and Yang \cite{Yang}.

\begin{figure}[tbp]
\includegraphics[width=3.2in]{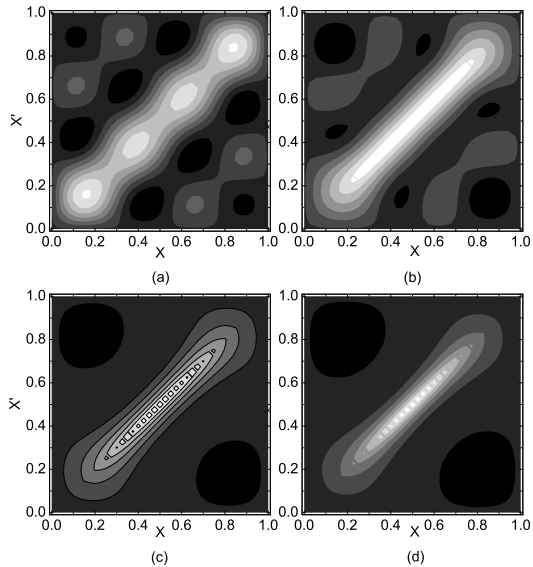}
\caption{ Gray scale plot of the one-body density matrix of Fermions
with $p$-wave attractive interactions as a function of the
dimensionless coordinates $x$ and $x'$ for $N=4$. (a) $c_F=0$; (b)
$c_F=-0.1$; (c) $c_F=-1$; (d) $ c_F=-\infty$.} \label{fig1}
\end{figure}

\begin{figure}[tbp]
\includegraphics[width=3.5in]{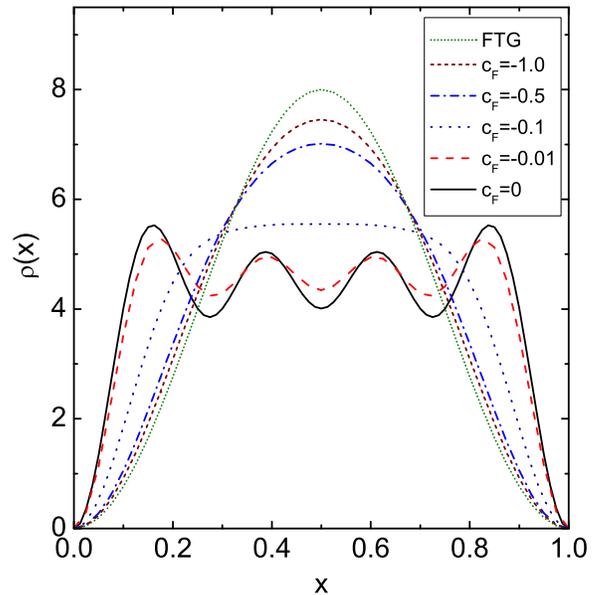}
\caption{(color online) Density distribution of Fermions with
$p$-wave attractive interactions for $N=4$.} \label{fig2}
\end{figure}

Before we proceed to the general case with intermediate
interaction strength, we'd like to take a look at the situation in
the two limiting cases. In the limit of free Fermion case, we have
$k_j=j\pi $ ($j=1,\cdots ,N$). In this exceptional case of
$c_F=0$, the quasi-momentum acquires the genuine physical meaning
of momentum and the atoms occupy the single particle momentum
states according to Pauli exclusion principle. We thus have the
total wave function
\begin{eqnarray*}
&&\Psi \left( x_1,x_2,\cdots ,x_N\right) \\
&=& C\sum_Q (-1)^Q\theta \left( x_{q_N}-x_{q_{N-1}}\right) \cdots
\theta \left( x_{q_2}-x_{q_1}\right)\\
&&\times \sum_P (-1)^P \prod_{j=1}^N\sin k_{p_j} x_{q_j},
\end{eqnarray*}
with $C$ the normalization constant, which is nothing but the Slater
determinant of the lowest $N$ eigenstates $\sin (j\pi x)$ of the
system of single particle in the hard wall potential. In the other
limit of strongly attractive interaction, i.e., the FTG limit, all
quasi-momenta take the same value $k_j=\pi $ ($j=1,\cdots ,N$) and
$\varphi_Q \left( x_{q_1},x_{q_2},\cdots ,x_{q_N}\right)
=C(-1)^Q\prod_{j=1}^N\sin \pi x_{q_j}$. Thus we have the total wave
function
\begin{eqnarray*}
&&\Psi \left( x_1,x_2,\cdots ,x_N\right) \\
&=&C\sum_Q(-1)^Q\theta \left( x_{q_N}-x_{q_{N-1}}\right) \cdots
\theta \left( x_{q_2}-x_{q_1}\right) \prod_{j=1}^N\sin \pi
x_{q_j}.
\end{eqnarray*}
On the other hand, the Fermi-Bose mapping method has been used to
give exactly the ground state of the FTG gas in \cite{Girardeau06}
\[
\Psi \left( x_1,x_2,\cdots ,x_N\right) =CA\left( x_1,x_2,\cdots ,x_N\right)
\prod_{j=1}^N\sin \pi x_j,
\]
with the antisymmetric function $A\left( x_1,x_2,\cdots
,x_N\right) =\prod_{1\leq j<l\leq N}$sgn$\left( x_l-x_j\right) $.
It is obvious that the above two wave functions match each other.

\section{ground state properties and comparison with Bosons}
We now turn to the system with finite interaction strength
$-\infty<c_F<0$. In this case the quasi-momenta are decided by
numerically solving the Bethe ansatz equations and the total wave
function is obtained by Eq.(\ref{WF}) through $\varphi
_Q(x_{q_1},x_{q_2},\cdots ,x_{q_N})$ under the restriction of
exchange antisymmetry. For the one dimensional interacting system, a
quantity of fundamental importance in many-body physics is the
one-body density matrix, which, in terms of the ground state wave
function $\Psi \left( x_1,\cdots ,x_N\right)$, is given by
\begin{eqnarray*}
&&\rho (x,x^{\prime }) \\
&=&\frac{N\int_0^Ldx_2\cdots dx_N\Psi ^{*}\left( x,x_2,\cdots ,x_N\right)
\Psi \left( x^{\prime },x_2,\cdots ,x_N\right) }{\int_0^Ldx_1\cdots
dx_N\left| \Psi \left( x_1,x_2,\cdots ,x_N\right) \right| ^2}.
\end{eqnarray*}
This quantity furnishes the expectation values of single particle
observables such as the position density distribution $\rho
(x)=\rho (x,x^{\prime })|_{x=x^{\prime }}$, and the momentum
distribution which is simply the Fourier transformation of $\rho
(x,x^{\prime })$,
\begin{equation}
n \left( k\right) =\frac 1{2\pi }\int_0^Ldx\int_0^Ldx^{\prime }\rho
(x,x^{\prime })e^{-ik\left( x-x^{\prime }\right) }.
\end{equation}

We display the one-body density matrix and the position density
distribution in Fig. 1 and in Fig. 2 for different $p$-wave
attractive interactions. The one-body density matrix expresses the
self correlation and it means the probability that two successive
measurements, one immediately following the other, will find the
particle at the point $x$ and $x^{\prime }$, respectively. We
notice that for all interacting strengthes there exists a strong
enhancement of the diagonal contribution $\rho (x,x^{\prime })$
along the line $x=x^{\prime }$. To see this more clearly the
position density distribution $\rho (x)$ for $N=4$ particles is
shown in Fig. 2 with more variable interacting strengthes. For
noninteracting system the atoms behave as ideal Fermions and the
density profiles show obvious spatial oscillation structure.
Increasing the strength of the attractive $p$-wave interaction, as
seen in Fig. 2, first leads to the depression of the amplitude of
the oscillations of the density profile, followed by the emergence
of the typical Gaussian-like bosonic behavior and contraction of
the half-width of the density. In the limit of infinitely strong
attractive interaction between the Fermions, the system enters
into the FTG regime, and the density shows the same smooth profile
as that of noninteracting Bosons. Particularly, the density
distribution in Fig. 2 for FTG is identical to that of the free
Bose gas and there exists one-to-one correspondence for the
density distribution between the attractive fermionic gas and the
repulsive bosonic gas with the interacting strength related by
$c_F=-1/c$ \cite{Hao}.

\begin{figure}[tbp]
\includegraphics[width=3.5in]{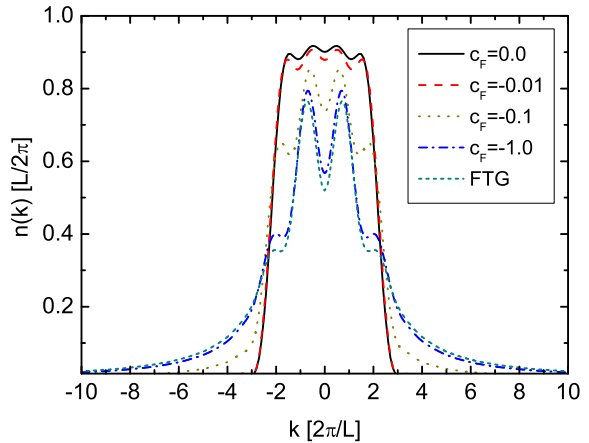}
\caption{(color online) Momentum distribution of Fermions with
$p$-wave attractive interactions for $N=4$.} \label{fig3}
\end{figure}

\begin{figure}[tbp]
\includegraphics[width=3.5in]{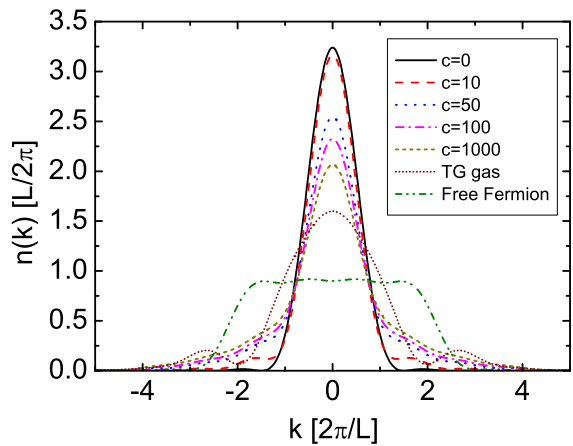}
\caption{(color online) Momentum distribution of Bosons with repulsive interactions for $%
N=4 $.} \label{fig5}
\end{figure}

\begin{figure}[tbp]
\includegraphics[width=3.5in]{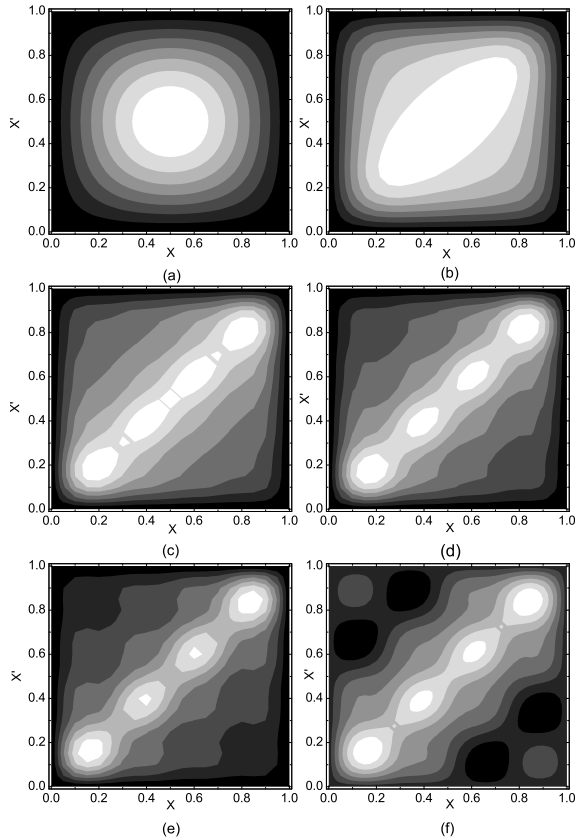}
\caption{ Gray scale plot of the one-body density matrix of Bosons
with repulsive interactions for $
N=4$. (a) $c=0$; (b) $c=10$; (c) $c=50$; (d) $c=100$; (e) $c=1000$; (f) $%
c=\infty $. } \label{fig4}
\end{figure}

It is worth to emphasize that the density distribution is identical
for TG bosons and noninteracting fermions, and also for FTG fermions
and noninteracting bosons. Nevertheless, their momentum
distributions show remarkable differences as shown in Fig. 3 for
fermions with different attractive interaction strengths and in Fig.
4 for bosons with the corresponding repulsive interaction strengths.
The momentum distribution for fermions oscillate in the full regime
and the number of oscillation peaks remains equal to the number of
atoms in the system. It becomes more and more nonuniform in the
momentum space and and two sharp spikes indicating that there is
high probability of finding the atom in momentum states around
$k\sim1$. Two other peaks at higher momentum diminish with
increasing interaction but remain there even in the FTG limit due to
the fermionic statistics. The half-width of the profiles become
larger as the attractive interaction increases.

As a comparison, the momentum distributions of Bosons with repulsive
interactions are given in Fig. 4 which are obtained by the Fourier
transformation of the one-body density matrix shown in Fig. 5.
Obviously, the momentum distribution for the bosons exhibits quite
different behaviors. As shown in Fig. 4, there is an obvious peak
around the zero momentum point and the height of the peak shrinks
with the increase of the repulsive interaction. Furthermore we find
no oscillation at all in the momentum distribution for bosons. Even
in the Tonks limit, the momentum distribution of Bosons does not
show shell structure like free Fermions. The largest probability of
distribution appears around the zero point of momentum and decreases
rapidly for finite momentum values. Stronger repulsive interaction
between the bosonic atoms tends to spread out the distribution into
higher momentum space. Although the momentum distribution for the
interacting bosonnic gases has been studied by different numerical
methods \cite{Astrakharchik,Deuretzbacher}, an exact result has
never been given except in the TG limit \cite{Lapeyre,Lenard}.

From the eqs. (\ref{Energy}) and (\ref{BAE}), we see that the
energy level structure of our fermionic model is exactly the same
as the corresponding bosonic model with interaction strength
related by $c=-1/c_F$. Therefore the thermodynamic properties of
the fermionic atoms with attractive $p$-wave interactions are the
same as the well known thermodynamic properties of the 1D boson
gas \cite{Gaudin,Lieb,Yang} with inverse coupling $c=-1/c_F$. This
implies that there exists a Bose-Fermi duality between the
$p$-wave fermionic model and bosonic one with arbitrary
interactions which can not be distinguished by the thermodynamic
properties. However, due to the different exchange symmetry of the
wave functions, the observables associated with the wave functions
rather than the square of wave functions (density distribution)
should display different behaviors. As we have shown, the
off-diagonal density matrix and the momentum distributions are
different greatly, which result from the different statistics
properties followed by Bosons and Fermions.

\section{Conclusions}
In summary, we have investigated the ground-state properties of
fermionic gases with attractive $p$-wave interactions in a
one-dimensional hard-wall trap. With the Bethe ansatz method, the
explicit wave function of the ground state and therefore the
one-body density and momentum distributions are obtained. It turns
out that the density distributions show one-to-one correspondence
between Fermions with attractive $p$-wave interactions and Bosons
with repulsive interactions. For weak attractive interaction the
density distributions of Fermions display shell structures and the
Boson-like distributions appear as the interactions become
stronger. In the FTG limit, the Fermi gas exhibits the same
distribution as that of the ideal Bose gas. This again confirms
the Bose-Fermi duality: strongly interacting Bosons behave like
Fermions, and vice versa. Nevertheless, from the viewpoint of
momentum distribution, the conclusion is rather different. In the
full interacting regime the momentum distributions of Fermions
show typical oscillations, which is in sharp contrast with the
case of bosonic atoms. In the Tonks limit of infinite interaction,
although the density profiles of one dimensional bosons display
the Fermion-like distribution, the momentum distribution is still
Boson-like.

\begin{acknowledgments}
S.C. is supported by NSF of China under Grant No. 10574150 and
programs of Chinese Academy of Sciences. Y.Z. is supported by 973
Program under Grant No. 2006CB921102 and Shanxi Province Youth
Science Foundation under Grant No. 20051001. The authors would like
to acknowledge the hospitality of KITPC in Beijing.
\end{acknowledgments}

\end{document}